\documentclass[11pt,showpacs,showkeys,aps,pra]{revtex4}
\usepackage[utf8x]{inputenc}
\usepackage{ucs}

\usepackage{amsmath,amssymb}
\usepackage{latexsym}
\usepackage{amsfonts}
\usepackage{dcolumn}% Align table columns on decimal point
\usepackage{bm}% bold math
\usepackage[usenames]{color}
\usepackage{multirow}
\usepackage{graphicx}
\usepackage{hyperref}
\usepackage{subfig}
\usepackage{booktabs}
\usepackage{xcolor}
\usepackage[normalem]{ulem}
\usepackage{float}
\usepackage{float} % Allows putting an [H] in \begin{figure} to specify the exact location of the figure
\usepackage{wrapfig} % Allows in-line images such as the example fish picture
\usepackage{upgreek} % para poner letras griegas sin cursiva.
\usepackage{cancel} % para tachar.
\usepackage{mathdots} % para el comando \iddots
\usepackage{mathrsfs} % para formato de letra
\usepackage{blindtext}
\usepackage{diagbox}
\usepackage{tabularx}
\usepackage{changepage}

\usepackage{lscape}
\usepackage{multirow}% http://ctan.org/pkg/multirow
\usepackage{hhline}% http://ctan.org/pkg/hhline
\usepackage{booktabs}% http://ctan.org/pkg/booktabs

\graphicspath{{figures/}} % Specifies the directory where pictures are stored

\renewcommand{\arraystretch}{2.5} 

\begin{document}

\title{Extremum-entropy-based Heisenberg-like uncertainty relations}

\author{I.V. Toranzo$^{a,c}$, S. L\'opez-Rosa$^{b,c}$, R.O. Esquivel$^{c,d}$, J.S. Dehesa$^{a,c}$}
\affiliation{
$^a$Departamento de F\'isica At\'omica, Molecular y Nuclear, Universidad de Granada, 18071-Granada, Spain\\
$^b$Departamento de F\'isica Aplicada II, Universidad de Sevilla , 41012-Sevilla, Spain\\
$^c$Instituto {\em Carlos I} de F\'isica Te\'orica y Computacional, Universidad de Granada, 18071-Granada, Spain\\
$^d$Departamento de Qu\'{\i}mica, Universidad Aut\'onoma Metropolitana, 09340-M\'exico D.F., M\'exico
}

\email{dehesa@ugr.es}
\date{\today}

\begin{abstract}
 In this  work we use the extremization method of various information-theoretic measures (Fisher information, Shannon entropy, Tsallis entropy) for $d$-dimensional quantum systems, which complementary describe the spreading of the quantum states of natural systems. Under some given constraints, usually one or two radial expectation values, this variational method allows us to determine an extremum-entropy distribution, which is is the \textit{least-biased} one to characterize the state among all those compatible with the known data. Then we use it, together with the spin-dependent uncertainty-like relations of Daubechies-Thakkar type, as a tool to obtain relationships between the  position and momentum radial expectation values of the type $\langle r^{\alpha}\rangle^{\frac{k}{\alpha}}\langle p^k\rangle\geq f(k,\alpha,q,N), q=2s+1,$ for $d$-dimensional systems of N fermions with spin s. The resulting uncertainty-like products, which take into account both spatial and spin degrees of freedom of the fermionic constituents of the system, are shown to often improve the best corresponding relationships existing in the literature.
\end{abstract}


\maketitle

\section{Introduction}

Let us consider a $d$-dimensional system of $N$ fermions with spin $s$ characterized by the wavefunction $\psi(\vec{r_{1}},\ldots,\vec{r_{N}};\sigma_{1},\ldots,\sigma_{N})$, for $\vec{r_{i}}\in \mathbb{R}^{d}$ and $\sigma_i \in \{1,2,\ldots,q\equiv 2s+1\}$, being antisymmetric in the pairs $(\vec{r_{i}},\sigma_{i})$ for all $i$. Then the norm $\langle \psi |\psi \rangle$ is 
\[
\langle \psi |\psi \rangle = \sum_{\sigma_{i}=1}^{q}\int |\psi(\vec{r_{1}},\ldots,\vec{r_{N}};\sigma_{1},\ldots,\sigma_{N})|^{2}\,\prod_{i=1}^{N}d^d r_{i},\quad i = 1,2,\ldots,N
\]
and the position single-particle density associated to the antisymmetric N-particle state $\psi$ is defined by
\begin{equation}
\label{eq:spd}
\rho(\vec{r}) = N\sum_{\sigma_{i}=1}^{q} \int |\psi(\vec{r},\ldots,\vec{r_{N}};\sigma_{1},\ldots,\sigma_{N})|^{2}\,\prod_{i=2}^{N}d^d r_{i},
\end{equation}
which is completely characterized by the knowledge of its radial expectation values $\langle r^{k}\rangle = \int_{\mathbb{R}_{d}} r^{k}\rho(\vec{r})\, d^{d}r$.
A similar statement can be written in momentum space for the momentum (i.e., Fourier-transformed) wavefunction $\tilde{\psi}(\vec{p_{1}},\ldots,\vec{p_{N}};\sigma_{1},\ldots,\sigma_{N})$, the momentum single-particle density $\gamma(\vec{p})$ and the momentum moments $\langle p^{k}\rangle = \int_{\mathbb{R}_{d}} p^{k}\gamma(\vec{p})\, d^{d}p$. Moreover, the notation $r = |\vec{r}|$, $p = |\vec{p}|$, and atomic units ($e= m =\hbar = 1$) have been used throughout the paper. 

The search for relationships (generally of inequality type) which interconnect properties of the position and momentum densities of quantum systems has been of permanent interest from the very beginning of quantum mechanics \cite{Heisenberg} up to now \cite{Lieb1,Lieb2,Parr-Yang,Sukumar,Lehtola,Thakkar1,Thakkar2, Thakkar-Pedersen, Lieb, Zozor} for both fundamental (e.g., mathematical realizations of the Heisenberg uncertainty principle, stability of matter) and applied (e.g., electronic structure of natural systems) reasons. These single-particle densities are completely characterized by the knowledge of the radial expectation values in the two conjugate spaces which often describe (aside of a numerical factor) numerous fundamental quantities of the system (see e.g., \cite{Gadre1986,Liu-parr3,Thakkar1}), for instance the magnetic susceptibility $\langle r^{2}\rangle$, the height peak of the Compton profile $\langle p^{-1}\rangle$ the kinetic energy $\langle p^{2}\rangle$, the relativistic Breit-Pauli energy $\langle p^{4}\rangle$, the total electron-electron repulsion energy $\langle p^{3}\rangle$, etc. Emphasis is usually centered about the inequality-type relations among the expectation values $\langle r^{n}\rangle$ of the position density $\rho(\vec{r})$ and the momentum radial expectation values $\langle p^{n}\rangle$ \cite{Thakkar-Blair1,Hart,Thakkar-Pedersen,Yue,Gadre1986,Tao1,Tao2} not only in three dimensions but also for quantum systems of arbitrary dimensionality $d$ \cite{Angulo1,dehesa4,dehesa5,dong}. All these inequality-type relations rely on some mathematical constraints on the momentum density $\gamma(\vec{p})$.\\

In this work, we have employed a novel procedure to obtain direct links between the expectation values $\langle r^{\alpha}\rangle$ and the momentum expectation values $\langle p^{k}\rangle$ for $d$-dimensional quantum systems, which starts with the inequality-type relationships between the momentum radial expectation values $\langle p^{k}\rangle$ and the entropic moments of the position density, $W_{n}[\rho]$ defined as
\begin{equation}
\label{eq:entropmom}
W_{n}[\rho]=\int [\rho(\vec{r})]^{n}\,d^{d}r.
\end{equation}
 The position entropic moments $W_{n}[\rho]$ describe, and/or are closely related to, some fundamental and/or experimentally accesible quantities (see e.g., \cite{Thakkar1,Thakkar-Pedersen}), such as e.g. Thomas-Fermi kinetic energy $W_{\frac{5}{3}}[\rho]$, the Dirac-Slater exchange energy $W_{\frac{4}{3}}[\rho]$, the Patterson function of x-ray crystallography $W_{3}[\rho]$, etc. In fact, the energetic quantities of the many-electron systems can be expressed in terms of these entropic moments as already pointed out \cite{Liu-parr1,Liu-parr2} in the framework of the density theory functional\cite{Parr-Yang}\\
Recently, it has been argued that the the momentum expectation values and the position entropic moments for $d$-dimensional systems of $N$ fermions with spin $s$ fulfil the following semiclassical spin-dependent uncertainty-like relations of Daubechies-Thakkar type \cite{Daubechies,Thakkar1,Thakkar-Pedersen} (see also \cite{tor}):
 \begin{equation}
 \label{eq:avevalmom4}
 \langle p^{k}\rangle \geq K_{d}(k)q^{-\frac{k}{d}} W_{1+\frac{k}{d}}[\rho],
 \end{equation}
 where $k>0$, and
 \begin{equation}
 \label{eq:Kdk}
  K_{d}(k) = \frac{d}{k+d}(2\pi)^{k}\frac{\left[\Gamma\left(1+\frac{d}{2}\right)\right]^{k/d}}{\pi^{k/2}}\, .
 \end{equation}
 Note that for $k<0$, the sign of inequality (\ref{eq:avevalmom4}) is inverted. Also note that these expressions simplify for three-dimensional systems as
 \begin{equation}
\label{eq:empiricavevalmom2}
\langle p^{k}\rangle \leq c_{k} W_{1+\frac{k}{3}}[\rho] \quad \text{for} \quad k=-2,-1
\end{equation}
and
\begin{equation}
\label{eq:empiricavevalmom3}
\langle p^{k}\rangle \geq c_{k} W_{1+\frac{k}{3}}[\rho] \quad \text{for} \quad k=1,2,3,4
\end{equation}
with $c_{k}=3(3\pi^{2})^{k/3}(k+3)^{-1}$ (since $K_{d}(k) = 2^{\frac{k}{3}}c_{k}$ for $d =3$ and $s=1/2$) , which were previously found by means of numerous semiclassical and Hartree-Fock-like ground-state calculations in atoms and diatomic molecules \cite{Pathak, Thakkar1, Hart, Thakkar-Pedersen, Porras1, Porras2}. Moreover, the case $k = 2$ in Eq. (\ref{eq:avevalmom4}) was previously conjectured by Lieb (see e.g. \cite{Lieb1}) and weaker versions of it have been rigorously proved, as discussed elsewhere \cite{Thakkar-Pedersen}. In fact, Eq. (\ref{eq:avevalmom4}) with constant $K'_{d}(k) = K_{d}(k) \times B(d,k)$ with $B(d,k) = \left\{\Gamma\left(\frac{d}{k}\right)\inf_{a>0}\left[ a^{-\frac{d}{k}}\left(\int_{a}^{\infty}du\, e^{-u}(u-a)u^{-1}\right)^{-1}\right] \right\}^{-\frac{k}{d}}$ has been rigorously proved by Daubechies \cite{Daubechies}. Let us also mention that a number of authors have published some rigorous $d$-dimensional bounds of the same type \cite{Lieb2, Hundertmark} with much less accuracy. Furthermore, let us point out that the inclusion of the spin $s$ in the lower bound (\ref{eq:avevalmom4}) for the expected value of $\langle p^2 \rangle$ was considered by Hundertmark \cite{Hundertmark} and applied to obtain uncertainty-like relations in \cite{Irene}; the extension to $\langle p^k \rangle$ has been recently used \cite{tor} in a similar sense.\\

The second step in our procedure is to invoke the information-theoretical grounds of Tao \cite{Tao0,Tao1,Tao2} and other authors \cite{Finkel,Gadre1986,dehesa3,angulo4,dehesa4,Romera3,sheila1,sheila2} to estimate to a very good approximation the position entropic moments $W_{n}[\rho]$ by means of the bounds provided by the entropic moments of the extremum-entropy distribution obtained with the extremum-entropy principle of information theory. This principle provides an interesting constructive method which objectively estimates the unknown distribution when only partial information (e.g., some radial expectation values) is given about a probability distribution. This method gives rise to an extremum-entropy distribution which is the ‘least-biased’ (minimally prejudiced) one among all those compatible with the known data, which are the constraints to be imposed in the variational problem when solving it by determining the values of the associated Lagrange multipliers. Indeed, for a generic information-theoretic measure $Q[\rho]$ subject to the constraints $\langle r^{0} \rangle \equiv \int \rho(\vec{r})\, d^{d}r = N$ and $\langle r^{\alpha} \rangle = \int r^{\alpha}\rho(\vec{r})\, d^{d}r$, $\alpha >0$, we extremize it by taking variations of the form
\[
\delta\left\{Q[\rho] - \lambda\int r^{\alpha}\rho(\vec{r})\, d^{d}r -\mu\int \rho(\vec{r})\, d^{d}r \right\}=0,
\]
where $\lambda$ and $\mu$ are Lagrange multipliers (see e.g. \cite{sheila1,sheila2}). The best known and most useful information-theoretic quantities, which complementary describe the spreading properties of the probability distribution $\rho(\vec{r})$, being $\vec{r} = (x_{1}, x_{2}, \ldots, x_{d})$, all over the space are the Shannon entropy $S[\rho]$ defined (see e.g. \cite{cover_91}) by 
\begin{equation}
\label{eq:Shannon}
S[\rho] := -\int \rho(\vec{r})\ln \rho(\vec{r})\, d^{d}r,
\end{equation}
the Tsallis entropy $T_{q}[\rho]$ given \cite{tsallis} by
\begin{equation}
\label{eq:Tsallis}
T_{t}[\rho]:= \frac{1}{t-1}\left\{1 - \int [\rho(\vec{r})]^{t}\, d^{d}r  \right\} \, ;\quad t>0,\, t\neq 1.
\end{equation}
and the Fisher information $I[\rho]$ of the density which is defined \cite{Frieden} by
\begin{equation}
\label{eq:Fisher}
I_{\rho} := \int \rho(\vec{r})\left( \frac{|\vec{\nabla}_{d}\rho(\vec{r})|}{\rho(\vec{r})}\right)^{2}\, d^{d}r\, ,
\end{equation}
(where $\vec{\nabla}_{d}$ denotes the $d$-dimensional gradient operator), respectively. Let us point out the well-known fact that for $t\rightarrow 1$ the Tsallis entropy $T_{q}$ is equal to the Shannon value $S_{\rho}$. The corresponding extremization problems associated to the Shannon and Tsallis entropies are the maximization entropy problems, briefly called as \textit{MaxEnt} and \textit{MaxTent} problems; and the one associated with the Fisher information is called by minimization Fisher problem (briefly, \textit{MinInf} problem); see e.g. \cite{sheila1,sheila2}) and references therein for further details.\\

Hereafter we use this two-step method to obtain the lower bounds to the Heisenberg-like products $\langle r^{\alpha}\rangle \, \langle p^k\rangle$ of $N$-fermion systems by use of the analytical solutions of the MaxEnt (see Section II), MinInf (see Section III) and MaxTent (see Section IV) problems with the above mentioned constraints ($\langle r^{0} \rangle = N, \langle r^{\alpha} \rangle$). Some particular cases are numerically examined for a large set of neutral atoms and, moreover, they are compared with the corresponding ones obtained by other authors.

\section{MaxEnt-based Heisenberg-like uncertainty relation}

Here we will apply the methodology described in the previous section to find the Heisenberg-like uncertainty products $\langle r^{\alpha}\rangle \, \langle p^k\rangle$ of $d$-dimensional $N$-fermion systems by use of the analytical solution $\rho_{S}(r)$ of the MaxEnt problem with the constraints ($\langle r^{0} \rangle = N, \langle r^{\alpha} \rangle$). Then, we center around the corresponding products in the three-dimensional case, and finally we consider a few particular cases. 

Following the lines described in \cite{sheila1,sheila2}, the $d$-dimensional density which maximizes the Shannon entropy (\ref{eq:Shannon}) when the constraints correspond to one radial expectation value $\langle r^{\alpha}\rangle$ in addition to the normalization to the number of particles $N$ is given by
\begin{equation}
\label{eq:maxentdens}
\rho_{S}(r) = e^{-\lambda -\mu \, r^{\alpha }}; \quad \alpha >0\, ,
\end{equation}
where the Lagrange multipliers have the form
\begin{eqnarray}
\label{eq:lambda0d}
\lambda &=& \log \left(\frac{2 \pi ^{d/2} d^{-\frac{d}{\alpha }} \alpha ^{\frac{d}{\alpha }-1} \langle r^{\alpha}\rangle^{d/\alpha } \Gamma \left(\frac{d}{\alpha }\right) }{\Gamma \left(\frac{d}{2}\right)}N^{-\frac{d}{\alpha }-1}\right)\\ 
\label{eq:lambda1d}
\mu &=& \frac{d \,N}{\alpha  \langle r^{\alpha}\rangle}.
\end{eqnarray}
Now, we compute the entropic moments (\ref{eq:entropmom}) of the maximizer solution (\ref{eq:maxentdens}) and insert them into the semiclassical position-momentum inequality (\ref{eq:avevalmom4}). We have found the following set of $d$-dimensional uncertainty-like relations:
\begin{equation}
\label{eq:URmaxentd}
\langle r^{\alpha}\rangle^{\frac{k}{\alpha}}\langle p^k\rangle \geq 2^{\frac{(d-2) k}{d}}d^{\frac{(\alpha +d) (d+k)}{\alpha  d}} (d+k)^{-\frac{\alpha +d}{\alpha }} \alpha ^{k \left(\frac{1}{d}-\frac{1}{\alpha }\right)} q^{-\frac{k}{d}} \Gamma \left(\frac{d}{2}\right)^{\frac{d+k}{d}} \Gamma \left(\frac{d}{\alpha }\right)^{-\frac{k}{d}} N^{k \left(\frac{1}{\alpha }+\frac{1}{d}\right)+1},
\end{equation}
with $\alpha > 0$, $k > 0$ and $q = 2s + 1$. It is straightforward to obtain that for real N-electron systems ($d=3$ and $q=2$), one has that the uncertainty-like relations simplify as
%\begin{equation}
%\label{eq:lambda03}
%\lambda_{3} = \log \left[ 4 \pi  3^{-3/\alpha } \alpha ^{\frac{3}{\alpha }-1} \Gamma \left(\frac{3}{\alpha }\right) \langle r^{\alpha}\rangle^{3/\alpha } N^{-\frac{3}{\alpha }-1}\right]
%\end{equation}
%and
%\begin{equation}
%\label{eq:lambda13}
%\mu_{3} = \frac{3\,N}{\alpha\langle r^{\alpha} \rangle},
%\end{equation}

\begin{equation}
\label{eq:URmaxent3}
\langle r^{\alpha}\rangle^{\frac{k}{\alpha}}\langle p^k\rangle \geq \frac{2^{-\frac{2 k}{3}} \pi ^{\frac{k}{3}} 3^{\frac{(\alpha +3) (k+3)}{3 \alpha }}   \Gamma \left(\frac{3}{\alpha }\right)^{-\frac{k}{3}}\alpha ^{k\frac{(\alpha -3)}{3 \alpha }}  }{(k+3)^{1+\frac{3}{\alpha}}}N^{k\left(\frac{1}{\alpha }+\frac{1}{3}\right)+1}.
\end{equation}
Some particular cases of these inequalities are shown in the Table \ref{table:table1}. Finally, let us remark that these results coincide with the corresponding ones obtained by Tao et al \cite{Tao0,Tao1,Tao2} with a similar methodology in a few particular three-dimensional cases derived one by one by these authors for the lowest orders of the radial expectation values, which constitutes a test of our results.. For completeness, we give in Table \ref{table:tableMAXTAO} the approximate values for the lower bounds of a few Heisenberg-like products. The accuracy of these uncertainty-like relations has also been studied for almost all the neutral atoms \cite{Tao1} and numerous diatomic molecules \cite{Tao2}. 
{\tiny
\begin{table}[H]
\begin{center}
\setlength{\tabcolsep}{3.0pt}
\renewcommand{\arraystretch}{1.2}% Tighter
\begin{tabular}{|c|c|c|c|c|}
\hline
\multicolumn{5}{|c|}{$\langle r^{\alpha}\rangle^{\frac{k}{\alpha}}\langle p^{k}\rangle\geq f\left(k,\alpha,N\right)$}\\
\hline
\diagbox{$k$}{$\alpha$} &$1$ & $2$ & $3$ & $4$\\
\hline
$1$  & $\frac{243}{512}(3\pi)^{1/3}N^{7/3}$ & $\frac{27\,3^{1/3}\pi^{1/6}}{32\sqrt{2}} N^{11/6}$ & $\frac{9}{16}\left(\frac{3}{2}\right)^{2/3}\pi^{1/3} N^{5/3}$ & $\frac{9}{16}\left(\frac{3\pi}{\Gamma(3/4)}\right)^{1/3} N^{19/12}$\\
\hline
$2$  & $\frac{729 (3 \pi )^{2/3} }{2500}N^{11/3}$ & $\frac{81 \sqrt[6]{3} \sqrt[3]{\pi } }{50 \sqrt{5}}N^{8/3}$ & $\frac{27}{50} \sqrt[3]{\frac{3}{2}} \pi ^{2/3} N^{7/3}$ & $\frac{9\ 3^{11/12} \pi^{2/3} }{10\ 5^{3/4}\left[\Gamma \left(\frac{3}{4}\right)\right]^{2/3}}N^{13/6}$\\
\hline
$3$  & $\frac{81 \pi }{128} N^5$ & $\frac{9}{16} \sqrt{3 \pi } N^{7/2}$ & $\frac{9 \pi  }{16}N^3$ & $\frac{3\ 3^{3/4} \pi  }{8 \sqrt[4]{2} \Gamma \left(\frac{3}{4}\right)}N^{11/4}$\\
\hline
$4$ & $\frac{19683 \sqrt[3]{3} \pi ^{4/3} }{38416}N^{19/3}$ & $\frac{243\ 3^{5/6} \pi ^{2/3} }{196 \sqrt{7} }N^{13/3}$ & $\frac{81}{196} \left(\frac{3}{2}\right)^{2/3} \pi ^{4/3} N^{11/3}$ & $\frac{81 \sqrt[12]{3} \,\pi^{4/3} }{28\ 7^{3/4}\left[\Gamma \left(\frac{3}{4}\right)\right]^{4/3}}N^{10/3}$\\
\hline
\end{tabular}
\end{center}
\caption{Some MaxEnt-based Heisenberg-like uncertainty relations for $N$-electron systems. The contribution of both spatial and spin degrees of freedom are taken into account.}
\label{table:table1}
\end{table}
}

{\tiny
\begin{table}[H]
\begin{center}
\setlength{\tabcolsep}{3.0pt}
\renewcommand{\arraystretch}{1.}% Tighter
\begin{tabular}{@{}*{5}{p{.2\textwidth}@{}}}
%\hline
%\multicolumn{2}{|c|}{$\langle r^{-1}\rangle^{-k}\langle p^{k}\rangle\geq f(N)$}\\
\hline
$\langle r\rangle^{3}\langle p^{3}\rangle $  & $\langle r^{2}\rangle^{\frac{3}{2}}\langle p^{3}\rangle $  & $\langle r^{3}\rangle \langle p^{3}\rangle $  &  $\langle r\rangle^{2}\langle p^{2}\rangle $   & $\langle r^{2}\rangle \langle p^{2}\rangle $ \\
$1.98804\,N^5$ & $1.72686\,N^{\frac{7}{2}}$ & $1.76715\,N^3$ & $1.30107\,N^{\frac{11}{3}}$  & $1.27429 \,N^{\frac{8}{3}}$ \\
\hline
$\langle r^{3}\rangle^{\frac{2}{3}}\langle p^{2}\rangle $  &  $\langle r\rangle \langle p\rangle $ & $\langle r^{2}\rangle^{\frac{1}{2}}\langle p\rangle $ &  $\langle r^{3}\rangle^{\frac{1}{3}}\langle p\rangle $ & \\
$1.32594 \,N^{\frac{7}{3}}$ & $1.00252 \,N^{\frac{7}{3}}$  & $1.04135\,N^{\frac{11}{6}}$ & $1.07953\, N^{\frac{5}{3}}$ & \\
\hline
\end{tabular}
\end{center}
\caption{Numerical values of the MaxEnt-based lower bounds for various Heisenberg-like uncertainty products $\langle r^{\alpha}\rangle^{\frac{k}{\alpha}}\langle p^{k}\rangle$.}
\label{table:tableMAXTAO}
\end{table}
}

\section{MinInf-based Heisenberg-like uncertainty relation}
In this section we apply the methodology to obtain the Heisenberg-like uncertainty products $\langle r^{-1}\rangle \, \langle p^k\rangle$ of $d$-dimensional systems of $N$ fermions with spin $s$, by use of the analytical solution $\rho_{I,d}(r)$ of the MinInf problem (i.e., minimization of the Fisher information) with the constraints ($\langle r^{0} \rangle = N, \langle r^{-1} \rangle$). Thereupon, the resulting expressions are applied to real systems of finite many-electron systems and, for illustrative purposes, the instance $\langle r^{-1}\rangle^{-2}\langle p^{2}\rangle$ is compared with the best corresponding result published in the literature.\\

 According to the lines developed in \cite{Romera3} (see also \cite{sheila1,sheila2}) we have determined the following expression for the minimizer density 
\begin{equation}
\label{eq:minfinfdensd}
\rho_I(r) = \frac{2^{-d}\pi ^{\frac{1-d}{2}} (d-1)^d}{\Gamma \left(\frac{d+1}{2}\right)}N^{1-d} \langle r^{-1} \rangle^de^{-\frac{(d-1) \langle r^{-1} \rangle}{N}\,r},
\end{equation}
which simplifies for the three dimensional case as 
\begin{equation}
\label{eq:minfinfdens3}
\rho_{I,3}(r) = \frac{1}{\pi  N^2}\langle r^{-1}\rangle^3 e^{-\frac{2 \langle r^{-1} \rangle }{N}r}.
\end{equation}
Then, the semiclassical position-momentum inequality (\ref{eq:avevalmom4}) together with the entropic moments of the minimizer density $\rho_{I,d}(r)$ allow us to obtain the uncertainty-like products 
\begin{equation}
\label{eq:urminfinfd}
\langle r^{-1}\rangle^{-k}\langle p^k\rangle \geq d^{d+1} (d-1)^k (d+k)^{-d-1} \pi ^{\frac{k}{2 d}}\left[\frac{\Gamma \left(\frac{d}{2}+1\right)}{\Gamma \left(\frac{d+1}{2}\right)}\right]^{k/d} q^{-\frac{k}{d}}N^{\left(\frac{1}{d}-1\right) k+1},
\end{equation}
with $k > 0$, and $q$ denotes the number of spin states of each constituent fermion as already mentioned. In the particular case where $d=3$ and $s=1/2$, this inequality reduces to the following uncertainty-like relation 
\begin{equation}
\label{eq:urminfinf3}
\langle r^{-1}\rangle^{-k}\langle p^k\rangle \geq \frac{3^{\frac{k}{3}+4} \pi ^{k/3} }{(k+3)^4}N^{1-\frac{2 k}{3}},
\end{equation}
which is valid for all antisymmetric wavefunctions of $N$-electron systems.\\

In Table \ref{table:table2} a few particular cases of this Heisenberg-like relation are given for the lowest values of $k$.
{\tiny
\begin{table}[H]
\begin{center}
\setlength{\tabcolsep}{3.0pt}
\renewcommand{\arraystretch}{1.5}% Tighter
%\begin{tabular}{|@{}*{5}{p{.2\textwidth}@{}}|}
\begin{tabular}{|c|c|c|c|c|}
\hline
\multicolumn{5}{|c|}{$\langle r^{-1}\rangle^{-k}\langle p^{k}\rangle\geq f\left(k,N\right)$}\\
\hline
$k$ & $1$ & $2$ & $3$ & $4$  \\
\hline
 $f(N)$ & $\frac{81}{256} \sqrt[3]{3 \pi } N^{1/3}$ & $\frac{81 (3 \pi )^{2/3}}{625} N^{-1/3}$ &  $\frac{3 \pi }{16}N^{-1}$ & $\frac{243 \sqrt[3]{3} \pi ^{4/3}}{2401}N^{-5/3}$ \\
\hline
\end{tabular}
\end{center}
\caption{Some Heisenberg-like uncertainty relations for $N$-electron systems obtained by use of the MinInf density. The contributions of both spatial and spin degrees of freedom have been taken into account.}
\label{table:table2}
\end{table}
}
Let us now compare these lower bounds on the Heisenberg-like products $\langle r^{-1}\rangle \, \langle p^k\rangle$ (obtained here by use of the MinInf density)  with the corresponding MaxEnt-based results obtained by Tao et al \cite{Tao1,Tao2}, which seems to be the best one published in the literature, at least for the instance $\langle r^{-1}\rangle^{-2}\langle p^{2}\rangle$. In particular, in Table \ref{table:tableMINFTAO}  we can appreciate that the MinInf-based lower bound on the uncertainty-like product $\langle r^{-1}\rangle^{-2}\langle p^{2}\rangle$ is a factor 1.25 better than the corresponding MaxEnt-based one \cite{Tao1,Tao2}. 
{\tiny
\begin{table}[H]
\begin{center}
\setlength{\tabcolsep}{4.0pt}
\renewcommand{\arraystretch}{1.}% Tighter
\begin{tabular}{|c|c|c|}
%\hline
%\multicolumn{2}{|c|}{$\langle r^{-1}\rangle^{-k}\langle p^{k}\rangle\geq f(N)$}\\
\hline
& MinInf-based bound & MaxEnt-based bound\\
\hline
$\langle r^{-1}\rangle^{-2}\langle p^{2}\rangle $ & $0.5783 N^{-\frac{1}{3}}$& $0.4615 N^{-\frac{1}{3}}$  \\
\hline
\end{tabular}
\end{center}
\caption{Comparison of the MinInf-based lower bounds obtained in the present work and the MaxEnt-based \cite{Tao1,Tao2} ones.}
\label{table:tableMINFTAO}
\end{table}
}
Finally, we should point out that there exist other lower bounds on the Heisenberg-like products $\langle r^{-1}\rangle \, \langle p^k\rangle$, but they only take into account the proper contribution of the spatial degrees of freedom. In particular, an interesting Rényi-based bound on the product $\langle r^{-1}\rangle^{-2}\langle p^{2}\rangle$, has been recently published \cite{Angulo3} , where the electron density has been averaged for spin and normalized to unity.  Hence, proper care has to be taken in comparing this with our results. The proper consideration of the contribution of the spin degrees of freedom decreases the lower bound, thus worsening its accuracy.

\section{MaxTent-based Heisenberg-like uncertainty relation}

Let us now use the previous methodology to determine the Heisenberg-like uncertainty products $\langle r^{\alpha}\rangle \, \langle p^k\rangle$ of three-dimensional $N$-fermion systems by use of the analytical solution $\rho_{T}(r)$ of the MaxTent problem (i.e., the maximization of the Tsallis entropy $T_{t}[\rho]$) with the constraints ($\langle r^{0} \rangle = N, \langle r^{\alpha} \rangle$).\\ 
According to the lines given in \cite{sheila1,sheila2}, we first determine the density $\rho_{_{T}}(r)$ which maximizes the Tsallis entropy (\ref{eq:Tsallis}) when the constraints are $(N, \langle r^{\alpha}\rangle)$. There are two different cases:
\begin{itemize}
\item If $t > 1$ and $\alpha> 0$, the maximum entropy density only exists for a finite interval $r \in [0, a]$, where $a$ is a parameter to be determined within the framework of our variational procedure, as a function of the corresponding Lagrange multipliers.

\item If $0<t<1$ and $\alpha >3\frac{1-t}{t}$, the maximum entropy density exists for any value of $r$.
\end{itemize}

\subsection{Case $0<t<1$, $\alpha>3\frac{1-t}{t}$ and $k<0$}
In this case the Tsallis maximizer density is 
\begin{equation}
\label{eq:maxtentdens1}
\rho_T(r) = C\left[\frac{1}{t}(a^{\alpha }+r^{\alpha })\right]^{\frac{1}{t-1}},
\end{equation}
where
\begin{eqnarray}
\label{eq:a1}
a &=& \left[\frac{\langle r^{\alpha}\rangle \alpha\,  \Gamma \left(\frac{1}{1-t}-\frac{3}{\alpha }\right)}{N 3\, \Gamma \left(-\frac{t}{t-1}-\frac{3}{\alpha }\right)}\right]^{1/\alpha } \\
\label{eq:C1}
C &=& t^{\frac{1}{t-1}} 3^{\frac{3}{\alpha }+\frac{1}{t-1}} \alpha ^{-\frac{3}{\alpha }+\frac{1}{1-t}+1}  \langle r^{\alpha}\rangle^{\frac{1}{1-t}-\frac{3}{\alpha }} N^{\frac{3}{\alpha }+\frac{1}{t-1}+1} \nonumber\\
& \times &\frac{\Gamma \left(\frac{1}{1-t}\right)\Gamma \left(\frac{1}{1-t}-\frac{3}{\alpha }\right)^{-\frac{3}{\alpha }-\frac{t}{t-1}} \Gamma \left(-\frac{t}{t-1}-\frac{3}{\alpha }\right)^{\frac{3}{\alpha }+\frac{1}{t-1}}}{4 \pi  \Gamma \left(\frac{3}{\alpha }\right)}
\end{eqnarray}
Now we compute the entropic moments (\ref{eq:entropmom}) of the maximizer solution (\ref{eq:maxtentdens1}) and insert them into the semiclassical position-momentum inequality (\ref{eq:avevalmom4}), obtaining the following set of uncertainty-like relations:
\begin{eqnarray}
\label{eq:TmaxUR1}
\langle r^{\alpha }\rangle^{\frac{k}{\alpha }}\langle p^k\rangle &\leq& 2^{-\frac{k}{3}}\pi^{\frac{k}{3}} 3^{\left(\frac{1}{\alpha }+\frac{1}{3}\right) k+1} \alpha ^{\frac{(\alpha -3) k}{3 \alpha }} q^{-\frac{k}{3}} N^{\left(\frac{1}{\alpha }+\frac{1}{3}\right) k+1} \nonumber\\
&\times& \frac{\Gamma \left(\frac{3}{\alpha }\right)^{-\frac{k}{3}} \Gamma \left(\frac{1}{1-t}\right)^{\frac{k}{3}+1} \Gamma \left(\frac{k+3}{3(1-t)}-\frac{3}{\alpha }\right) \Gamma \left(\frac{1}{1-t}-\frac{3}{\alpha }\right)^{-\left(\frac{1}{\alpha }+\frac{1}{3}\right) k-1} \Gamma \left(\frac{t}{1-t}-\frac{3}{\alpha }\right)^{k/\alpha }}{(k+3) \Gamma \left(\frac{k+3}{3(1- t)}\right)},\nonumber\\
\end{eqnarray}
with $\alpha>3\frac{1-t}{t}$, $0<t<1$ and $k<0$, which are valid for all antisymmetric states of real systems of $N$ fermions with spin $s$. \\
In the particular case where $k=-1$, $\alpha = 2$ and $s=1/2$ one obtains with $t \geq 0.78$ the uncertainty-like product
\begin{equation}
\label{eq:r2pm1}
\langle r^{2}\rangle\langle p^{-1}\rangle^{-2}\leq \frac{2^{2} \pi^{\frac{1}{3}}3^{-\frac{1}{3}} ( t-1)\Gamma \left(\frac{1}{1-t}-\frac{3}{2}\right)^{4/3} \Gamma \left(\frac{2}{3(1-t)}\right)^2}{(3 - 5 t) \Gamma \left(\frac{2}{3(1-t)}-\frac{3}{2}\right)^2 \Gamma \left(\frac{1}{1-t}\right)^{\frac{4}{3}}}N^{-\frac{1}{3}},
\end{equation}
which is valid for all antisymmetric states of N-electron systems. A similar expresssion can be derived for an upper bound on the product $\langle p^{2}\rangle \langle r^{-1}\rangle^{-2}$. Their interest lies in the fact that the former uncertainty product allows one to correlate the diamagnetic susceptibility and the peak of the Compton profile (which are equal, except for a factor, to $\langle r^{2}\rangle$ and $\langle p^{-1}\rangle$, respectively, as previously mentioned), whilst the latter involves the kinetic and electron-nucleus attraction energy (which, except for a factor, are equal to $\langle p^{2}\rangle$ and $\langle r^{-1}\rangle$, respectively, as already mentioned).\\
Moreover, we have 
\begin{equation}
\label{eq:r2pm1t0.78}
0.4368 \, N^{-\frac{1}{3}} \leq \langle r^{2}\rangle\langle p^{-1}\rangle^{-2}\leq 0.4958\, N^{-\frac{1}{3}},
\end{equation}
where we have used the upper bound (\ref{eq:r2pm1}) with $t=0.78$ and the lower bound $(12N)^{-\frac{1}{3}}$ recently obtained (see Eq. ($22$) of \cite{tor}). A similar chain of inequalities can be written for the product $\langle r^{-1}\rangle^{-2}\langle p^{2}\rangle$.

\subsection{Case $t>1$, $\alpha>0$ and $k>0$}

In this case the Tsallis maximizer density is 
\begin{equation}
\label{eq:maxtentdens2}
\rho_T(r) = C\left[\frac{1}{t}(a^{\alpha }-r^{\alpha })\right]^{\frac{1}{t-1}},
\end{equation}
where 
\begin{eqnarray}
\label{eq:a2}
a &=& \left(\frac{\langle r^{\alpha}\rangle (\alpha  t+3 (t-1))}{N (3 (t-1))}\right)^{1/\alpha }\\
\label{eq:C2}
C &=& \frac{\alpha\,  t^{\frac{1}{t-1}} \left(\frac{\langle r^{\alpha}\rangle (\alpha  t+3 (t-1))}{N (3 (t-1))}\right)^{-\frac{3}{\alpha }-\frac{1}{t-1}}}{4 \pi  B\left(\frac{3}{\alpha },\frac{t}{t-1}\right)},
\end{eqnarray}
where $B(x,y) = \frac{\Gamma(x) \Gamma(y)}{\Gamma(x+y)}$. Then, the calculation of the entropic moments (\ref{eq:entropmom}) of the maximizer solution (\ref{eq:maxtentdens2}) and the semiclassical position-momentum inequality (\ref{eq:avevalmom4}) allows us to determine the following set of uncertainty-like relations:
\begin{eqnarray}
\label{eq:TmaxUR2}
\langle r^{\alpha }\rangle^{\frac{k}{\alpha }}\langle p^k\rangle &\geq& q^{-\frac{k}{3}} N^{\left(\frac{1}{\alpha }+\frac{1}{3}\right) k+1}\left(\frac{\pi }{2}\right)^{k/3} 3^{\left(\frac{1}{\alpha }+\frac{1}{3}\right) k+1} \alpha ^{k/3}  \left(\frac{\alpha  t}{t-1}+3\right)^{-\frac{k}{\alpha }}(k+3)^{-1} \nonumber\\
& & \times\frac{\Gamma \left(\frac{k+3 t}{3 (t-1)}\right) \Gamma \left(\frac{3}{\alpha }\right)}{\Gamma \left(\frac{k+3 t}{3 (t-1)}+\frac{3}{\alpha }\right)B\left(\frac{3}{\alpha },\frac{t}{t-1}\right)^{1+\frac{k}{3}}},
\end{eqnarray}
with $\alpha>0$, $k> 0$  and $t>1$, which are valid for all antisymmetric states of real $N$ fermions with spin $s$. A few particular cases of this set obtained with $t=2$ are given in Table \ref{table:table4}.

{\tiny
\begin{table}[H]
\begin{center}
\setlength{\tabcolsep}{7.0pt}
\renewcommand{\arraystretch}{1.}% Tighter
\begin{tabular}{|c|c|c|c|}
\hline
\multicolumn{4}{|c|}{$\langle r^{\alpha}\rangle^{\frac{k}{\alpha}}\langle p^{k}\rangle\geq f\left(k,\alpha,q,N\right)$}\\
\hline
\diagbox{$k$}{$\alpha$} & $1$ & $2$ & $3$ \\
\hline
$1$  & $1.22995 N^{7/3}q^{-\frac{1}{3}}$ & $1.30485 N^{11/6}q^{-\frac{1}{3}}$ & $1.35791 N^{5/3}q^{-\frac{1}{3}}$ \\
\hline
$2$  & $1.67378 N^{11/3}q^{-2/3}$ & $1.8647 N^{8/3}q^{-2/3}$ & $2.00783 N^{7/3}q^{-2/3}$ \\
\hline
$3$  & $2.4429 N^5q^{-1}$ & $2.83315 N^{7/2}q^{-1}$ & $3.14159 N^3q^{-1}$ \\
\hline
\end{tabular}
\end{center}
\caption{Some MaxTent-based Heisenberg-like uncertainty relations for real systems of $N$-fermions, where the contribution of both spatial and spin degrees of freedom are taken into account.}
\label{table:table4}
\end{table}
}
To illustrate the accuracy and validity of these Heisenberg-like uncertainty relations the MaxTent-based lower bounds for two Heisenberg-like products are compared in Table \ref{table:tableMAXTAO3} with the corresponding MaxEnt-based ones of Tao et al \cite{Tao1,Tao2} (also obtained in Section II of this work in a generic and unified way) and the corresponding Hartree-Fock values (obtained by means of the accurate near-Hartree-Fock wavefunctions of Koga et al \cite{koga1,koga2}) for all neutral atoms from He to Xe. Remark that the MaxTent-based lower bounds obtained in the present work are systematically higher (that is, better) than the corresponding bounds of Tao et al. Moreover, we observe that there is still a large gap between our bounds and the Hartree-Fock values to be fulfilled. This gap clearly grows when the atomic number increases. Nevertheless we should keep in mind that the Hartree-Fock values of the Heisenberg-like products are atom dependent, while the corresponding uncertainty products provided by our two-step method have an universal character in the sense that they are valid for any N-fermion system with arbitrary spin.
\begin{table}[H]
\begin{center}
\setlength{\tabcolsep}{7.0pt}
\renewcommand{\arraystretch}{1.3}% Tighter
\resizebox{\columnwidth}{!}{
\begin{tabular}{|c|c|c|c|c|c|c|c|}
\hline
%\multicolumn{2}{|c|}{$\langle r^{-1}\rangle^{-k}\langle p^{k}\rangle\geq f(N)$}\\
 \multicolumn{2}{|c|}{}  &\multicolumn{3}{|c|}{$\langle r\rangle\langle p\rangle$}  &\multicolumn{3}{|c|}{$\langle r^{2}\rangle^{\frac{1}{2}}\langle p\rangle$}\\
\hline
Atom  & $N$ & Tao et al & Present work & Hartree-Fock & Tao et al & Present work& Hartree-Fock  \\
\hline
He	& 2	& 5.052&	5.056	& 5.191 & 3.711	& 3.714& 4.309	 \\ \hline
Li	& 3&	13.013	& 13.022  & 24.626  &	7.804&	7.811	&  21.175\\ \hline
Be&	4&	25.462 &	25.481  & 45.563 &	13.225 &	13.235	& 30.938\\ \hline
B&	5&	42.856 &	42.888   & 72.529 &	19.9010 &	19.925	& 42.399\\ \hline
C&	6 &	65.580 &	65.629  & 103.326 &	27.812 &	27.834	&  53.709\\ \hline
N	& 7&	93.968 &	94.038	 & 138.645 & 36.895	& 36.924 & 65.568	\\ \hline
O	& 8	& 128.320 &	128.415  & 180.515&	47.128&	47.165&	 79.275 \\ \hline
Ne&	10&	215.982 &	216.143	 & 277.741& 70.950&	71.006	& 107.750 \\ \hline
Na	& 11&	269.774 &	269.975	 &  441.672 & 84.496	& 84.563 & 212.256\\ \hline
Mg	& 12&	330.502 &	330.748  & 570.342 &	99.110 &	99.188	& 253.193 \\ \hline
Al	& 13&	398.369 &	398.665  & 723.183 &	114.775&	114.865& 305.055 \\ \hline
Si	& 14&	473.569&	473.921  & 858.352 &	131.478&	131.581& 336.696 \\ \hline
P	& 15&	556.285 &	556.699  & 990.945 &	149.206&	149.323& 364.153 \\ \hline
S	& 16&	646.692&	647.173  & 1135.067 &	167.946&	168.079& 396.063 \\ \hline
Cl	& 17&	744.958&	745.512  & 1279.150 &	187.690&	187.837& 424.992 \\ \hline
Ar	& 18&	851.243 &	851.876  & 1425.453 &	208.425&	208.589& 452.573 \\ \hline
Ti	& 22&	1359.578 &	1360.589	 & 2633.898 & 301.110&	301.347& 863.813 \\ \hline
Fe	& 26	& 2007.654 &	2009.147  & 3557.697 &	409.011 &	409.333	& 1043.880  \\ \hline
Ni	& 28&	2386.637	& 2388.412	 & 4067.283 & 468.533&	468.902	& 1139.181 \\ \hline
Zn	& 30&	2803.501&	2805.585	 & 4606.755& 531.708& 	532.126	&  1238.371 \\ \hline
Ge	& 32&	3259.125&	3261.548  & 5628.555 &	598.493	& 598.964& 1500.495 \\ \hline
Se	& 34&	3754.354	& 3757.146	 & 6517.395 & 668.850	& 669.376&	1646.220 \\ \hline
Kr	& 36&	4289.996&	4293.186	 & 7383.097 & 742.743&	743.328	& 1769.071 \\ \hline
Xe	& 54&	11049.344&	11057.561  & 21127.164 &	1561.970&	1563.199& 4281.013 \\ \hline
\hline
\end{tabular}
}
\end{center}
\caption{Comparison of the MaxTent-based lower bounds on two Heisenberg-like products obtained in the present work, the MaxEnt-based ones of Tao et al and the corresponding Hartree-Fock values. The former ones were calculated with the Tsallis optimal parameter $t=3$ and $2.3$ for the two uncertainty products considered in this Table, respectively. }
\label{table:tableMAXTAO3}
\end{table}
%\end{landscape}

\section{Conclusions}

We have developed a two-step method to derive position-momentum Heisenberg-like uncertainty relations of the type $\langle r^{\alpha}\rangle^{\frac{k}{\alpha}}\langle p^k\rangle\geq f\left(k,\alpha,q,N\right), q=2s+1,$ for $d$-dimensional systems of $N$ fermions with spin $s$ which have two relevant features: they take into account the contribution of both the spatial and spin degrees of freedom, and they allow one to often improve the corresponding relationships published in the literature. First, the method makes use of the semiclassical Daubechies-Thakkar uncertainty-like inequality (\ref{eq:avevalmom4}) which connects the radial momentum expectation values of order $k, k>0$, and the position entropic moments of order $1 + \frac{k}{d}$. Then, the latter position density functionals are estimated by means of the corresponding functionals of the extremum-distribution density of the systems obtained via the extremization principle of various information-theoretic quantities (Shannon entropy, Fisher information, Tsallis entropy) subject to two constraints, the normalization of the single-particle density and a position radial expectation value of given order. \\

In summary, the resulting MaxEnt-based and MinInf-based uncertainty-like inequalities given by Eqs. (\ref{eq:URmaxentd}) and (\ref{eq:urminfinfd}), respectively, not only extend to $d$-dimensional systems the corresponding ones previously obtained in the literature for three-dimensional systems by use of different methodologies, but they often improve them. Moreover, the MaxTent-based inequality-like relationships given by Eqs. (\ref{eq:TmaxUR1}) and (\ref{eq:TmaxUR2}), here obtained for three-dimensional systems, seem to have in general a better accuracy than the corresponding MaxEnt-based ones (which are the most accurate values reported so far), as it is numerically shown here for various particular cases applied to a large variety of neutral atoms from He to Xe for illustrative purposes.\\

 Nevertheless, there is still large room for improving these inequalities. Much work along the lines of the present study is needed, not only because of the insufficient accuracy of these uncertainty-like relations but also because of the the correlation between fundamental and/or experimentally accessible quantities (e.g., the volume, the ionization energy, electronegativity, hardness and other atomic and molecular properties) is very relevant in the framework of the density functional theory, as already pointed out elsewhere \cite{Thakkar-Blair1,Thakkar1, Thakkar2,Blair,Thakkar-Pedersen,Hart,Lehtola,Tao2}.\\
 
Finally, we are aware that there exist other information-theoretic approaches to the uncertainty principle such as for example the application of majorization theory \cite{partovi,puchala,luis,bosyk,portesi,rudnicki}, which relies on the partial order on probability vectors to characterize uncertainty. Majorization-based formulations of the uncertainty principle, although mathematically more complex, complement the entropic and variance-based formulations, leading to a deeper knowledge of the fundamental aspects of uncertainty and disorder in quantum theory. They have been developed for finite-dimensional quantum systems, showing a great interest in the theory of quantum information. However, its extension to infinite-dimensional Hilbert spaces is presently very challenging. We wonder whether our work might provide hints towards the generalization of the majorization-like ideas for continuous variables. This is an open problem which deserves to be separately considered.

\section*{Acknowledgments}
 This work was partially supported by the Projects FQM-2445 and FQM-207 of the Junta de Andalucía and the MINECO grants FIS2011-24540, FIS2014-54497P and FIS2014-59311-P and the Mexican grant CONACYT: CB-2009-01/132224, as well as by FEDER. The work of I. V. Toranzo was supported by the program FPU of MINECO.

\end{document}